\documentstyle[aps,prd,psfig]{revtex}
\def\be{\begin{equation}}
\def\en{\end{equation}}

\begin{document}

\title{Relativistic Hydrodynamics around Black Holes
and Horizon Adapted Coordinate Systems}

\author{Philippos Papadopoulos and Jos\'e A. Font}

\medskip

\address{Max-Planck-Institut f\"ur Gravitationsphysik \\
Schlaatzweg 1, D-14473, Potsdam, Germany}

\medskip

\maketitle

\begin{abstract}

Despite the fact that the Schwarzschild and Kerr solutions for the
Einstein equations, when written in standard Schwarzschild and
Boyer-Lindquist coordinates, present coordinate singularities, all
numerical studies of accretion flows onto collapsed objects have been
widely using them over the years. This approach introduces conceptual
and practical complications in places where a smooth solution should
be guaranteed, i.e., at the gravitational radius.  In the present
paper, we propose an alternative way of solving the general
relativistic hydrodynamic equations in background (fixed) black hole
spacetimes. We identify classes of coordinates in which the (possibly
rotating) black hole metric is free of coordinate singularities at the
horizon, independent of time, and admits a spacelike decomposition. In
the spherically symmetric, non-rotating case, we re-derive exact
solutions for dust and perfect fluid accretion in
Eddington-Finkelstein coordinates, and compare with numerical
hydrodynamic integrations. We perform representative axisymmetric
computations. These demonstrations suggest that the use of those
coordinate systems carries significant improvements over the standard
approach, especially for higher dimensional studies.

\end{abstract}


\section{Introduction}
\label{sec:introduction}

It is well known that the black hole solutions of the field equations
of general relativity, with the Schwarzschild and Kerr solutions being
astrophysically the most relevant, exhibit coordinate singularities
when written in coordinates adapted to the exterior region. The very
notion of a black hole was greatly clarified with the discovery, in
the early sixties, of coordinate systems that remove those
singularities, and indeed cover the whole spacetime~\cite{Kruskal60}
(for a general overview see~\cite{MTW73},\cite{Hawking}). The
(rotating) black hole solution for the metric tensor, expressed in
standard Boyer-Lindquist coordinates $(t,r,\theta,\phi)$ is singular
at the roots of the equation $\Delta=r^2-2Mr+a^2=0$ (where $M$ is the
mass and $a$ the angular momentum aspect of the hole~\cite{units}). In
the spherically symmetric Schwarzschild case, the singularity at
$r=2M$ is removable with the use of appropriate transformations of the
radial and time coordinates. Slightly more complicated
transformations, involving also the azimuthal coordinate, can remove
the singularities at $r=r_{\pm}=M
\pm (M^2-a^2)^{1/2}$ of a rotating black hole.

The location of the coordinate singularity coincides with the event
horizon of the black hole. Asymptotic observers lose causal contact
with events near the black hole at precisely this location. Hence,
despite the singular appearance of the metric, in simulations of
matter flows in the background gravitational field of a black hole, one
is in principle allowed to consider the open interval extending from
the event horizon to some ``far zone" at a large, but finite, distance
from the hole.  This approach has been widely used over the years in
the numerical simulations of flows around black holes~\cite{Wi72}
-\cite{Font98a}.

The blow-up of the metric components at the horizon has implications
on the behavior of hydrodynamical quantities. The coordinate flow
velocity becomes ultra-relativistic and reaches the speed of light at
the horizon. As a consequence, the Lorentz factor becomes infinite
causing any numerical code to crash. Placing the inner boundary close
to the horizon, required for capturing the effects of the relativistic
potential, introduces large gradients in all hydrodynamical
variables. The steep radial behavior makes the task of numerical
evolution with a reasonable degree of accuracy challenging.

Besides practical considerations, the approach of evolving only the
exterior domain lacks of a well defined location for the inner
boundary: the computation cannot include the horizon surface, but must
commence at a location ``sufficiently close'' to it. Physically, the
influence of the horizon region on the solution will progressively
red-shift away the closer one gets to the horizon. The validity of a
certain choice though, must be continuously reasserted, as new flows
or effects are being investigated, with careful tests of convergence,
as the inner boundary is progressively moved inwards. Such tests are
complicated by the fact that, as mentioned in the previous paragraph,
the solutions appear singular at the horizon and hence demand
increasingly more resolution.

Ameliorating those problems has motivated the use of a logarithmic
radial coordinate (the so-called {\it tortoise}
coordinate~\cite{RW57}). This technique relegates the event horizon to
a infinite, negative, coordinate distance.  An equidistant grid in the
tortoise $r_{*}$ coordinate maps into an increasingly dense grid in
the Schwarzschild $r$ coordinate, with infinite density at the
horizon. This approach has proven successful in the extensive
semi-analytic studies of black hole perturbation theory, and recently
also for the axisymmetric integration of curvature perturbations as an
initial value problem~\cite{Krivanetal}. In wave systems, the
ambiguity of the location of the inner boundary is addressed by the
simple limiting form of the governing equations near the horizon. This
is the well known fact that black holes act as finite potential
barriers to electromagnetic and gravitational
perturbations~\cite{chandra}.

The use of a tortoise coordinate does not bring similarly extensive
benefits to the study of the hydrodynamical equations. The issue of
the inner boundary location is less transparent for those equations.
The steepness of the solution and the ``artificially'' high coordinate
velocities persist, although they are now more treatable due to the
substantial increase in resolution. In three dimensional simulations
using Cartesian coordinates, the tortoise technique is not possible at
all. It has been shown recently~\cite{Papadopoulos98a} that {\em
adaptive mesh refinement} can indeed provide, even for 3D systems, the
required resolution close to the event horizon. However, as alluded
to above, these undesired pathologies can be eliminated in the case of
fixed given black hole backgrounds with rather simple coordinate
transformations. This reserves the power of adaptive mesh refinement
for the more physically interesting features of the solution.

There is considerable freedom in choosing coordinates regular at the
horizon, which can be productively reduced by imposing criteria that
can enhance their suitability for numerical applications. The obvious
first criterion is of course the {\em regularity of the metric}, in
particular at the horizon. A second condition is that the constant
time surfaces are {\em everywhere spacelike}, as this is, currently, a
pre-requisite for the implementation of modern numerical methods for
relativistic hydrodynamic flows. An important third criterion is the
{\em time-independence} of the metric components. This leads to
constant in time coefficients in the equations and simplifies
disentangling the true hydrodynamical evolution from coordinate
effects in the black hole background. Interestingly, we will see that
those conditions still do not fix the coordinate system uniquely. We
show that the number of available choices greatly reduces, but is
still infinite, in the rotating black hole case. We give several
examples of such systems, which we collectively call {\em horizon
adapted coordinate systems}.

Such coordinate systems address the issues raised above in a
straight-forward way: any radius {\em inside} the horizon is equally
appropriate (in the idealized continuum limit) for the imposition of a
boundary condition, as the domain is causally disconnected from the
exterior. Importantly, the irrelevance of the inner boundary location
will persist even after the inclusion of other possible local physical
processes that may be considered in conjuction with the hydrodynamical
flow, e.g., radiative processes.  The coordinate velocities of the
flow will be bounded at the horizon, as they represent projections of
the (finite) fluid four velocity onto a regular coordinate
system. Hence the hydrodynamical nature of the flow becomes
considerably less demanding on the integration algorithm.  Gradients
in the solution for {\em scalar variables} such as pressure and
enthalpy will of course persist. Those are physical and due to the
curvature of the black hole, which requires a significant dynamic
range for its resolution.

The organization of this paper is as follows. In section II we
introduce a class of horizon adapted coordinate systems for a
non-rotating black hole. The rotating black hole and the more
restricted class of coordinates available in that case are discussed
in the Appendix.  We outline the numerical hydrodynamical framework of
our computations in section III. Exact solutions for spherical (Bondi)
accretion are presented in section IV for both dust and perfect
fluids. Section V describes the numerical results.  In our numerical
simulations we focus mainly on the spherically symmetric case, which
captures the essential nature of the problem. Some axisymmetric
computations are also briefly considered. The coordinate system on
which we base our computations is the celebrated Eddington-Finkelstein
coordinate system~\cite{Eddington24},\cite{Finkel58}.

Our main aim in this report is to show the functionality of this class
of coordinate systems as frameworks for the integration of the
equations of relativistic hydrodynamics in black hole spacetimes.  In
three-dimensions (and rotating holes) computations of accretion onto
black holes are likely to benefit significantly by the adoption of a
horizon adapted coordinate system.

\section{Horizon adapted coordinate systems}
\label{sec:hacs}

A transformation of the Schwarzschild time $t$ coordinate
to a new ``generic'' coordinate $\hat{t}$ according to
\begin{equation}
d\hat{t}=dt \pm \frac{1}{A(r)} 
\left(1- G(r) A(r)\right)^{1/2} dr ,
\label{eq:coord1}
\end{equation}
where $A(r) = 1-2M/r$, and $G(r)$ an arbitrary function, brings 
the Schwarzschild metric into the form
\begin{equation}
ds^2 = - A(r) d\hat{t}^2 
      \pm 2 \left(1 - G(r) A(r)\right)^{1/2} d\hat{t}dr 
       + G(r) dr^2 +
       r^2 (d\theta^2+\sin^2\theta d\phi^2) .
\end{equation}

The four-dimensional metric is next split following the $\{3+1\}$
framework, as
\begin{equation}
ds^2 = - (\alpha d\hat{t})^2 + \gamma_{ij} (dx^{i} + \beta^{i} d\hat{t})
(dx^{j} + \beta^{j} d\hat{t})
\end{equation}
and hence into the lapse function $\alpha=G(r)^{-1/2}$, the shift
vector $\beta_i = (\pm \left(1 - G(r) A(r)\right)^{1/2}, 0,0)$ or
$\beta^i=\gamma^{ij}\beta_{j}$, and the intrinsic spatial metric of
the constant $\hat{t}$ hypersurfaces 
$\gamma_{ij} = \mbox{diag}(G(r),r^2,r^2
\sin^{2}\theta)$.

We proceed to establish the features we stated above as defining the
horizon adapted coordinate systems. The spacelike nature of the
foliation is evident for any positive definite function $G(r)$, the
normal to the hypersurface given by $n^{\mu}=(G(r)^{1/2},\mp
G(r)^{-1/2} (1-G(r)A(r))^{1/2},0,0)$.  The independence of all metric
functions from the time coordinate $\hat{t}$ is also manifest.  The
regularity of the metric is also guaranteed if $G(r)$ is bounded from
above (we allow exceptions at the true curvature singularities).
Hence a large class of horizon adapted coordinates is parametrized by
the positive bounded functions $G(r)$. The essential action of the
coordinate transformation~(\ref{eq:coord1}) is to introduce a
non-diagonal $g_{tr}$ term (a radial shift vector). The shift covector
will always have $\beta_{r}=\pm 1$ at the horizon and the time flow
vector ($t^{\mu} =
\alpha n^{\mu} + \beta^{\mu}$) will be lightlike there.

The choice of sign in the above formulae affects whether it is the
past (-) or future (+) component of the event horizon that is
regularized with the coordinate transformations.  Some obvious choices
for $G(r)$ are $G(r) = 1$, $G(r) = (1 + 2M/r)$, $G(r) =
(1+2M/r)(1+(2M/r)^2)$, etc. The second choice leads to the well known
Eddington-Finkelstein (EF) coordinates which we will adopt for the
numerical work in this paper.  The coordinate
transformation~(\ref{eq:coord1}) and the relevant discussion of
regularity is easily generalized for a rotating black hole.  As shown
in the Appendix, in that case additional conditions of regularity on
an off-diagonal spatial metric component considerably reduce the class
of horizon adapted coordinates.  Further discussion of special cases
is given there. Simulations of accretion flows onto rotating black
holes in horizon adapted coordinates, more specifically in the
Kerr-Schild coordinate system, will be presented elsewhere~\cite{FontIP}.  We
concentrate now on the EF form of the metric, regularizing the future
component of the event horizon, which we will use for our
hydrodynamical evolutions.  With $G(r) = (1 + 2M/r)$ the metric reads
\begin{equation}
ds^2 = -\left(1-\frac{2M}{r}\right) d\hat{t}^2 +
        \frac{4M}{r} d\hat{t}dr +
        \left(1+\frac{2M}{r}\right) dr^2 +
       r^2(d\theta^2+\sin^2\theta d\phi^2),
\label{efmetric}
\end{equation} 
and correspondingly $\alpha=(\frac{r}{r+2M})^{1/2}$, 
$\beta_r = \left(2M/r\right)$ and 
$\gamma_{ij} = \mbox{diag}(1+2M/r,r^2,r^2 \sin^{2}\theta) $.

The EF form of the metric represents an analytic extension of the
Schwarzschild solution from the region $2M<r<\infty$ to $0<r<\infty$.
It is among the algebraically simplest choices and generalizes to the
case of a rotating black hole. In three dimensions it is naturally
expressed in pseudo-Cartesian coordinates (Kerr-Shild form).  An
important coordinate property of the EF system is its adaptation to
the null cone structure of the black hole: the ingoing radial null
geodesics are described in a particularly simple way.

Recent work clearly indicates that coordinate systems regular at the
horizon, adapted to the local light cone structure, carry considerable
advantages for the evolution of generic black holes.  More
specifically, the EF system and its generalization for {\em dynamic}
spherically symmetric spacetimes has been used, for the integration of
a self gravitating scalar field in interaction with a black
hole~\cite{mc96}. In three dimensional integrations of the Einstein
equations, the stable evolution of an isolated black hole has been an
important target in the field of numerical relativity. The EF
coordinates have been identified~\cite{Allen} as a point of departure
for such integrations of the Cauchy problem.  Recently, the Binary
Black Hole Grand Challenge Alliance project demonstrated~\cite{BBH1}
the first advection of a black hole in a three-dimensional
computational grid using the EF framework in boosted form. More
remarkably, ``eternal'' evolutions have been achieved~\cite{BBH2}
using a formulation of the Einstein equations based on ingoing
characteristic foliations.

Those features of the EF family of coordinate systems do not appear
particularly relevant for hydrodynamical computations, since generic
fluid flows are not tied to the spacetime null-cone structure in any
essential way. It is apparent though, that in simulations accounting
for the bidirectional {\em coupling} of matter to the spacetime
geometry, computations of black hole interactions with matter would
benefit significantly from this framework. This motivation breaks
the degeneracy among coordinate systems for our purposes.

\section{Relativistic hydrodynamics and Numerical method}

To demonstrate the feasibility of the procedure for numerical
studies, we compute the hydrodynamic accretion of fluid flows onto
black holes in the EF coordinate system. To do this, we solve the
general relativistic hydrodynamic equations in these particular
coordinates.  Recently, Banyuls et al.~\cite{Banyuls97} have written
these equations as an explicit hyperbolic system of conservation laws
(with sources) for a general metric written in $\{3+1\}$ form. Here,
we have to specialize those general expressions to the metric given
by Eq.~(\ref{efmetric}). The evolved quantities are the relativistic
densities of particle number, $D$, momentum (in the $i$ direction),
$S_i$ and energy, $\tau$. These quantities are defined in terms of a
set of {\it primitive} variables $(\rho,v_i,\epsilon)$ as $D=\rho W$,
$S_i=\rho h W^2 v_i$ and $\tau=\rho h W^2 - p - D$. Here, $\rho$ is
the rest-mass density, $p$ is the pressure, $v_i$ is the 3-velocity of
the fluid and $h$ is the specific enthalpy, defined as
$h=1+\epsilon+p/\rho$, with $\epsilon$ being the specific internal
energy.

The contravariant components of the 3-velocity are defined as
\begin{equation}
v^i=\frac{u^i}{\alpha u^t} +
\frac{\beta^i}{\alpha} ,
\end{equation}
where $u^{\mu}$ is the fluid 4-velocity, $\alpha$ is the lapse function
($\alpha^2=-1/g^{tt}$). Indices for
spatial vectors (the shift vector and the 3-velocity) are raised and
lowered using the spatial part of the metric, $\gamma_{ij}$. The
quantity $\alpha u^t$ is the Lorentz factor, $W$, which satisfies
$W=(1-v^2)^{-1/2}$ with $v^2=\gamma_{ij}v^i v^j$.

We have written a numerical code to solve the general relativistic
hydrodynamic equations in the background spacetime of a non-rotating
hole, using the EF coordinates. For this purpose we have taken
advantage of the fact that the equations form a hyperbolic system and
are explicitely written in conservation form~\cite{Banyuls97}.  This
allows for the use of advanced numerical methods based on approximate
(linearized) Riemann solvers whose improved ability to handle
relativistic (and ultrarelativistic) flows, high Mach number flows and
discontinuous solutions (shock waves), have been recently
investigated (see, e.g,~\cite{Banyuls97},\cite{Font98a},
\cite{MMFIM96} and references therein). The
Riemann solver used (either Roe~\cite{Roe81} or Marquina~\cite{dm},
\cite{detal}) relies on the spectral decomposition of the Jacobian
matrices of the multidimensional system of relativistic equations. The
spatial accuracy of the code is improved by means of a monotonic
piecewise-linear reconstruction~\cite{VL79} of proper rest-mass
density, flow velocity and internal energy. Integration in time is
done using a total-variation-diminishing Runge-Kutta scheme of
high-order, developed by Shu and Osher~\cite{Shu88}
We refer the interested reader to~\cite{Banyuls97} and~\cite{FIMM92}
for further details on the system of equations and
the numerical code.

\section{Exact solutions}

The feasibility of the procedure is shown with a comparison of the
numerical integration with exact solutions for spherically symmetric
accretion of dust and perfect fluid.  
Here, we re-derive those exact solutions in the EF coordinate system.

Following Michel~\cite{Michel72} we integrate the steady-state hydrodynamic
equations to obtain
\begin{equation}
\sqrt{-g}\rho u^r = C_1 , 
\label{eqc1}
\end{equation}
\begin{equation}
\sqrt{-g}\rho u^r u_t = C_2 ,
\label{eqc2}
\end{equation}
\noindent
where $C_1$ and $C_2$ are integration constants and $g$ stands for the
determinant of the four-metric, $\sqrt{-g}=\alpha\sqrt{\gamma}$, with
$\gamma=\det{(\gamma_{ij})}$. Hence, $u_t=C_2/C_1=C_3$. The 
coordinate velocity $v^r$ is determined by $C_3$ and $u_r$ through
the condition $g_{\mu\nu}u^{\mu}u^{\nu}=-1$. 

For the dust case, 
the final expressions are simple algebraic functions.
Fixing $u_t=C_3=-1$ (the marginally bound case) the solution reads
\begin{equation}
\rho (r)=\frac{-C_1}{r^2\sqrt{\frac{2M}{r}}} ,
\label{dens}
\end{equation}
\begin{equation}
v^r (r) = -\frac{1}{\sqrt{1+\frac{r}{2M}} (1+\sqrt{\frac{2M}{r}} + 
       \frac{2M}{r}) } ,
\end{equation}
\noindent
which is valid in the domain $0 < r < \infty$, i.e., also inside the
horizon. The radial dependence of the rest of the hydrodynamical quantities
can be directly computed from these two equations.

Let us turn now to the perfect fluid case. Monotonic steady-state
infall of perfect fluid onto a black hole must pass through the
critical point of the so-called ``wind'' equation, obtained by
differentiation of the integrated steady state equations
(Eqs.~(\ref{eqc1})~and~(\ref{eqc2})) with respect to $\rho$, $u^r$ and $r$, 
and elimination of the $d\rho$ differential,
\begin{equation}
\frac{du^r}{u^r}\left[ V^2 - \frac{(u^{r})^2}{u_{t}^2}\right]
+\frac{dr}{r}\left[2V^2-\frac{M}{ru_{t}^2}\right] = 0,
\label{wind}
\end{equation}
where $V^2 \equiv \frac{d\ln(\rho h)}{d\ln\rho} - 1$ and
$u_{t}^2=(u^{r})^2 - g_{tt}$.  Using the critical condition, that is,
the assumption that both brackets in Eq.~(\ref{wind}) vanish
simultaneously, the relation $(u^{r}_{c})^{2}=M/2r_{c}$ determines
$u^r_{c}$ and $(u_{t})_{c}$, after the selection of the critical point
radius $r_{c}$. From $u^r$ we get $u_t$ and hence we obtain $V^2$ at
the critical radius, $V_c^2=(u^r_c/u_{t_{c}})^2$. On the other hand,
considering a polytropic equation of state $P=K \rho^{\Gamma}$, 
with adiabatic index $\Gamma$ and a critical density
$\rho_{c}$, we can infer the value of $V_c^2$, which, in conjuction
with $V_c^2=(u^r_c/u_{t_{c}})^2$ determines the polytropic index $K$.
This complete determination of the fluid parameters at the critical
point, combined with the steady-state conditions, fixes the
integration constants $C_{1}$ and $C_{2}$, and hence uniquely
identifies the extension of the solution away from the critical
point. Contrary to the dust case, the solution does not have a closed
form, as one must solve a non-linear algebraic equation for $u^r(r)$,
from which one can get the rest of the variables. This is done using a
Newton-Raphson iteration.

\section{Results}

We now use the exact solutions derived in the previous section to
check the accuracy of our numerical evolutions and to demonstrate the
feasibility of our approach.  For the numerical investigation we
consider homogeneous initial data across the domain of integration.
The initial data evolve in time, until a final steady-state solution
is reached.  We focus mainly on the spherically symmetric case,
to which the exact solutions apply. At the end of the section
we describe some axisymmetric computations.

In spherical symmetry, the numerical domain extends from any given
non-zero radius inside the horizon, $r_{min}$, to some $r_{max}$
located in the physical universe (outside the horizon).  In Figs. 1
and 2 we plot a comparison of the exact and numerical solutions for
the dust and perfect fluid cases, respectively.  We show the final
steady-state for the spherical accretion problem. The solid lines
represent the exact solution and the circles indicate the numerical
one. In the particular computations considered here we have chosen
$r_{min}=0.5M$ and $r_{max}=50M$. We have used a non-uniform numerical
grid of 200 zones (logarithmic spacing). In both the dust and perfect
fluid integrations we have only assumed the exact solution in the
outermost zone. The integration has been performed until the final
steady-state solution has been achieved in the whole domain.  The
perfect fluid equation of state is given by $p=(\Gamma-1)\rho\epsilon$, 
with $\Gamma$ being the (constant) adiabatic exponent of the gas. 
We have chosen $\Gamma=4/3$.

The solutions are seen to be regular well inside the horizon at
$r=2M$. Nothing special happens at the horizon where all quantities
are smooth.  The steepness of the hydrodynamic quantities dominates
the solution only near the real singularity.  It is worth to stress
the behavior of the velocity in this new coordinate system, which
differs considerably from the Schwarzschild case. Now the total
velocity does not approach the speed of light as it reaches the
horizon, and although the maximum fluid speed is still at $r=2M$, it
is now considerably smaller, $v=\sqrt{\gamma_{rr}}v^r=1/\sqrt{3}$, in
the dust case. For the perfect fluid case this value is even smaller,
due to the pressure support of the gas. This is a most attractive
numerical behavior, as it eliminates the divergence in the Lorentz
factor at the horizon which would cause any relativistic fluid
dynamics code to crash.  As can be deduced from both figures we have
found very good agreement with the exact solution.  In
Fig.~(\ref{fig:rel_error}) we display the relative errors for density,
pressure and total velocity in the case of perfect fluid accretion.

The resolution benefits from the use of horizon adapted coordinates
appear to be considerable, hence we attempt a rough quantitative
assessment.  We perform a sequence of comparison integrations in
Schwarzschild and EF coordinates. We focus on the case of spherical
dust accretion. We use a grid of 100 zones and evolve the stationary
solution up to a final time of $100M$.  The outer boundary of the
domain is fixed at $50M$. The inner boundary of the EF domain is fixed
at $1.5M$. For the Schwarzschild computation we use a tortoise radial
coordinate which provides very high resolution near the horizon. We
have found that the solution in Schwarzschild coordinates is
maintained stationary, provided $r_{min}\ge 2.1M$. For $r_{min}=2.05M$
and a maximum resolution of $10^{-3}M$, the accretion departs from the
exact initial solution at the zones closer to the horizon. For
$r_{min}=2.01$ and same resolution, the code crashes. In this case, a
grid of 200 points does not preserve the initial solution either (even
with a resolution of $10^{-4}M$). At least 400 zones are required
(with a maximum $\Delta r \approx 10^{-5}$) to achieve the correct
accretion pattern.  On the other hand, using EF coordinates, we can
resolve the problem with an (unequally spaced) grid of 50 zones. This
gives, broadly speaking, a factor of 8 benefit in resolution when
using horizon adapted coordinate systems. Computations that may
require small values of $r_{min}$ will further increase this number,
the same being true for perfect fluid computations.  In this latter
case the computation in standard coordinates is more challenging, as
the pressure (or the internal energy) can easily become unstable near
the horizon and even become negative. Finally, three dimensional
Cartesian computations do not allow the use of tortoise coordinate
based grids. Hence a {\em conservative} estimate introduces an $8^3$
gain factor in the size of tractable problems in three dimensional
integrations.

We now briefly describe axisymmetric computations. We study the
so-called Bondi-Hoyle accretion, that is, hydrodynamic accretion onto
a moving black hole, in the EF coordinate system (for a
general overview of such flows see, e.g.,~\cite{Font98a} and
references therein). The asymptotic values of fluid velocity,
$v_{\infty}$, sound speed, $c_{s_{\infty}}$ and adiabatic exponent,
$\Gamma$, together with the assumption of an initially homogeneous
distribution suffice to define the initial values of all
hydrodynamical quantities. As in the spherical case, these data evolve
towards a final steady-state solution.  We use a grid of $200 \times
100$ zones in $r$ and $\theta$, respectively, with $r_{min}=1.5M$,
$r_{max}=100M$ and $0\le\theta\le\pi$. The radial grid is
logarithmically spaced. We evolve a model with $c_{s_{\infty}}=0.1$,
and $v_{\infty}=0.5$. Hence, it is a supersonic model with Mach number
$5$. We plot in Fig. 3 the stationary solution for $\Gamma=4/3$ (top),
$5/3$ (middle) and $2$ (bottom). We plot iso-contours of the logarithm
of the density. The most prominent feature of the solution is the
stationary conical shock. The inner circle indicates the position of
the horizon. Again, as in the spherical case, the solution smoothly
straddles the horizon at $2M$. Compared to the evolutions presented
in~\cite{Font98a} in Schwarzschild coordinates, we now use less
resolution to get to the same final solution. Highly supersonic models
with large values of $\Gamma$, in particular $\Gamma=5/3$, were very
hard to evolve with Schwarzschild coordinates and the $\Gamma=2$ case
was impossible to evolve. Its computation is now straightforward.

\section{Conclusions}

We have presented a family of {\it horizon adapted coordinate systems}
for the numerical study of accretion flows around black holes. In the
rotating case we identify a discrete but infinite family, of which the
first simple members are well known stationary coordinate systems. In
the non-rotating case the freedom in building the regular stationary
foliation is quantified by (at least) the space of bounded positive
functions of one variable.  We have shown how these systems allow for
a better numerical treatment of accretion scenarios. Existing numerical
studies of accretion flows onto black holes have been performed in the
original, singular systems, i.e., Schwarzschild coordinates for the
Schwarzschild solution and Boyer-Lindquist coordinates for the
rotating (Kerr) solution. Although it is possible to solve the problem
in these pathological coordinates, one is introducing artificial
complexity, being forced to use very high resolution to deal with the
unphysically large gradients that develop in the vicinity of the
horizon. This may prevent the accurate solution of three dimensional
problems.  At the same time, the ambiguity regarding the position of
the inner boundary of the domain (which should be the horizon)
introduces a convergence criterion that must be enforced at all times
if the solutions are to be trusted.

We have focussed on the particular case of the Eddington-Finkelstein
form of the Schwarzschild metric. The general relativistic
hydrodynamic equations are now regular at the horizon, which permits
an accurate description of accretion flows. We have demonstrated the
feasibility of this approach with the numerical study of the spherical
accretion (Bondi accretion) of dust and perfect fluid, and the 
comparison with the exact solutions which we re-derived in this
coordinate system. We have also shown the functionality of the new
coordinates in axisymmetric computations of relativistic
Bondi-Hoyle accretion flows.

In a forthcoming paper we plan to extend this approach to the rotating
case, considering horizon adapted coordinate systems to study
accretion flows in stationary Kerr spacetimes. Three dimensional
accretion flows onto black holes are interesting both from an
astrophysical and geometrical point of view, as they are thought to
correspond to observable electromagnetic emission, and hence may help
map the relativistic rotating black hole potential. The framework
proposed here will help detailed numerical studies of such systems in
the near future.

\section*{Acknowledgements}
We thank C.~Gundlach, J.M$^{\underline{\mbox{a}}}$~Ib\'a\~nez,
P.~Laguna and E.~Seidel for helpful discussions. P.P would like to
thank P.Anninos for his hospitality at NCSA where parts of this work were
completed. J.A.F acknowledges a fellowship from the TMR program of the
European Union (contract number ERBFMBICT971902). Computations were
performed on the SGI Origin2000 of the MPI f\"ur Gravitationsphysik at
Potsdam.

\appendix

\section*{}
\label{appendix}

In this appendix we present stationary coordinate systems for the Kerr
spacetime that are regular at and outside the horizon $r_{+}$, i.e.,
the larger of the roots of the equation $\Delta\equiv r^2-2Mr+a^2=0$.
In the Boyer-Lindquist form for the metric tensor the $\gamma_{rr}$
component is singular at $r_{+}$,
\begin{eqnarray}
ds^2 & = & -\frac{\Delta}{\rho^2} \left[ dt - 
a \sin^2\theta d\phi \right]^2 + \frac{\sin^2\theta}{\rho^2}
\left[(r^2 + a^2) d\phi - a dt \right]^2 \\  \nonumber
& & + \frac{\rho^2}{\Delta} dr^2 + \rho^2 d\theta^2 ,
\end{eqnarray}
with $\rho^2\equiv r^2 + a^2\cos^2\theta$.
We now introduce a coordinate transformation defined by
\begin{eqnarray}
d\hat{\phi} & = & d\phi - \frac{a}{\Delta} \epsilon(r,\theta) dr , \\
d\hat{t} &=  & dt - \epsilon(r,\theta)
\left[\frac{r^2+a^2}{\Delta} - \frac{\rho^2}{\rho^2 - 2Mr}
\left( 1 - \sqrt{1 - \left(1- \frac{2Mr}{\rho^2}\right) G(r,\theta)} 
\right) \right] dr , 
\label{kerrcoords}
\end{eqnarray}
where $\epsilon$ and $G$ are, at present, arbitrary functions.
This form of the transformation is chosen in
order to isolate the singular behavior of the spatial metric.
With the above ansatz, the metric becomes, in the new coordinates
$(\hat{t},r,\theta,\hat{\phi})$,
\begin{eqnarray}
ds^2 & = & - \left(1 - \frac{2Mr}{\rho^2}\right) d\hat{t}^2 
- \frac{4 M a r}{\rho^2} \sin^2\theta d\hat{t} d\hat{\phi} \\ \nonumber
& & - 2 \epsilon \sqrt{1-\left(1-\frac{2Mr}{\rho^2}\right)G}d\hat{t} dr
 + \left[ \frac{1-\epsilon^2}{\Delta} \rho^2 + G \right] dr^2 \\ \nonumber
& & + 2  \epsilon \frac{a \sin^2\theta \rho^2}{\rho^2-2Mr}
\left[ 1 - \frac{2Mr}{\rho^2} 
\sqrt{1 - \left(1- \frac{2Mr}{\rho^2}\right) G}
\right] d\hat{\phi} dr \\ \nonumber
& & + \rho^2 d\theta^2 + \sin^2\theta \left[
\rho^2 + a^2 \sin^2\theta \left(1 + \frac{2Mr}{\rho^2}\right) \right]
d\hat{\phi}^2.
\end{eqnarray}

By inspection of the spatial metric components we see that the
condition of regularity at the $\Delta=0$ surface demands that
$\epsilon^2 = 1$. This fixes the function $\epsilon$ up to a sign.  We
proceed to further examine the regularity of the metric element.  All
components of the metric, except $\gamma_{r\hat{\phi}}$, are regular,
except at the points $\rho^2=0$, which belong to the true singularity.
The component $\gamma_{r\hat{\phi}}$ is singular at the ergosphere
($\rho^2-2Mr=0$) which is {\em outside} the $r_{+}$ horizon, and coincides
with it at the poles $\theta=0,\pi$. 

Before we ameliorate this pathology we note that in the case of no
rotation $(a=0)$, the off-diagonal term drops, hence as we have
already seen in section~(\ref{sec:hacs}) there is no restriction on the
choice of the function $G$. This special behavior of the Schwarzschild
black hole allows selecting $G$ with the target of simplifying the
spatial 3-metric.  An interesting coordinate system, is obtained with
the simple choice $G(r)=1$, which brings the metric into the form
\begin{equation}
ds^2 = -\left(1-\frac{2M}{r}\right) d\hat{t}^2 +
       2\sqrt{\frac{2M}{r}} d\hat{t}dr +
       dr^2 + r^2(d\theta^2+\sin^2\theta d\phi^2),
\label{flat}
\end{equation}
and correspondingly $\alpha=1$, $\beta_r = \left(2M/r\right)^{1/2}$
and $\gamma_{ij} = \mbox{diag}(1,r^2,r^2 \sin^{2}\theta)$. This is the
Lemaitre coordinate form of the Schwarzschild
solution~\cite{Lemaitre}. The trivial form of the lapse and the
three-metric imply that all of the curvature information of the
black-hole spacetime is encoded in the way the $r,\theta,\phi$
coordinates must propagate from time slice to time slice (choice of
shift vector), in order to preserve a flat 3-space, and a constant
rate of clock ticking. It is quite remarkable that such a shift vector
is a simple, time-independent, function of radius, which, incidentally,
is equal to the {\em Newtonian} free fall velocity as measured locally
by static observers. This form of the metric element was the starting
point for a program of constructing black hole initial
data~\cite{Eardley}. Unfortunately, this simple state of affairs does
not appear generalizable to the rotating case.

Returning to the general case, by further exercising the freedom in
choosing the function $G(r,\theta)$ we can ensure that the
off-diagonal metric component is regular.

Substituting for $G$ via
\begin{equation}
1-\left(1-\frac{2Mr}{\rho^2}\right) G = Z^2 ,
\end{equation}
we see that the class of regular systems is parametrized
by the functions
\begin{equation}
Z_{k} = \left(\frac{2Mr}{\rho^2}\right)^{k} ,
\end{equation}
where $k$ is a non-negative integer.
The general admissible form for the function $G(r,\theta)$ is 
\begin{equation}
G_{k} = 1 + \frac{2Mr}{\rho^2} + \ldots + 
\left(\frac{2Mr}{\rho^2}\right)^{2k-1} ,
\end{equation}
for non-zero $k$, and 
\begin{equation}
G_{0} = 0 .
\end{equation}

With those substitutions the metric becomes
\begin{eqnarray}
ds^2 & = & - \left(1 - \frac{2Mr}{\rho^2}\right) d\hat{t}^2 
- \frac{4 M a r}{\rho^2} \sin^2\theta d\hat{t} d\hat{\phi} \\ \nonumber
& & +  \rho^2 d\theta^2 + \sin^2\theta \left[
\rho^2 + a^2 \sin^2\theta \left(1 + \frac{2Mr}{\rho^2}\right) \right]
d\hat{\phi}^2 \\ \nonumber
& & - 2 \epsilon Z_{k} d\hat{t} dr
 +  G_{k} dr^2  + 2  \epsilon a \sin^2\theta W_{k} d\hat{\phi} dr ,
\end{eqnarray}
where 
\begin{equation}
W_{k} = 1 + \frac{2Mr}{\rho^2} + \ldots + 
\left(\frac{2Mr}{\rho^2}\right)^{k} , 
\end{equation}
and all terms are regular except at the true curvature singularity.

{\em The case $k=0$} corresponds to the ingoing and outgoing
Eddington-Finkelstein coordinates ($\epsilon=-1$ and $\epsilon=1$
respectively). In this case the coefficient of the $\gamma_{rr}$
vanishes. Despite a common misconception, the foliation is still 
spacelike~\cite{Pretorius} for non-zero $a$. The null character
of the foliation in the limit $a=0$, though, implies that this
coordinate system is not appropriate for $3+1$ based studies.

{\em The case $k=1$} generates the Kerr-Schild coordinate
system, expressed in spherical coordinates via the 
relations~\cite{Hawking}
\begin{eqnarray}
x + i y & = & (r + i a) \sin\theta e^{i \hat{\phi}} , \\
z & = & r \cos\theta.
\end{eqnarray}
It is the algebraically simplest member of the family and reduces
to the EF system in the $a=0$ case.

{\em The case $k=2$} introduces a time coordinate that satisfies a
{\em harmonic} condition~\cite{Masso}, i.e., $\Box \hat{t} = 0$.  Some
recent hyperbolic formulations of the Einstein equations are based on
harmonic slicings, hence the black hole metric expressed in such
coordinate systems can be used effectively as a numerical code
test-bed~\cite{Sheel}.

The higher order members of the family, i.e., $k=3,4,\ldots$ may
also have interesting geometrical or differential properties.

\bibliographystyle{unsrt}

\begin{thebibliography}{20}
\bibitem{units} Geometrized
units $G=c=1$ are used throughout the paper.
\end{thebibliography}

\newpage

% FIGURE fig1.ps

\begin{figure}
\centerline{\psfig{figure=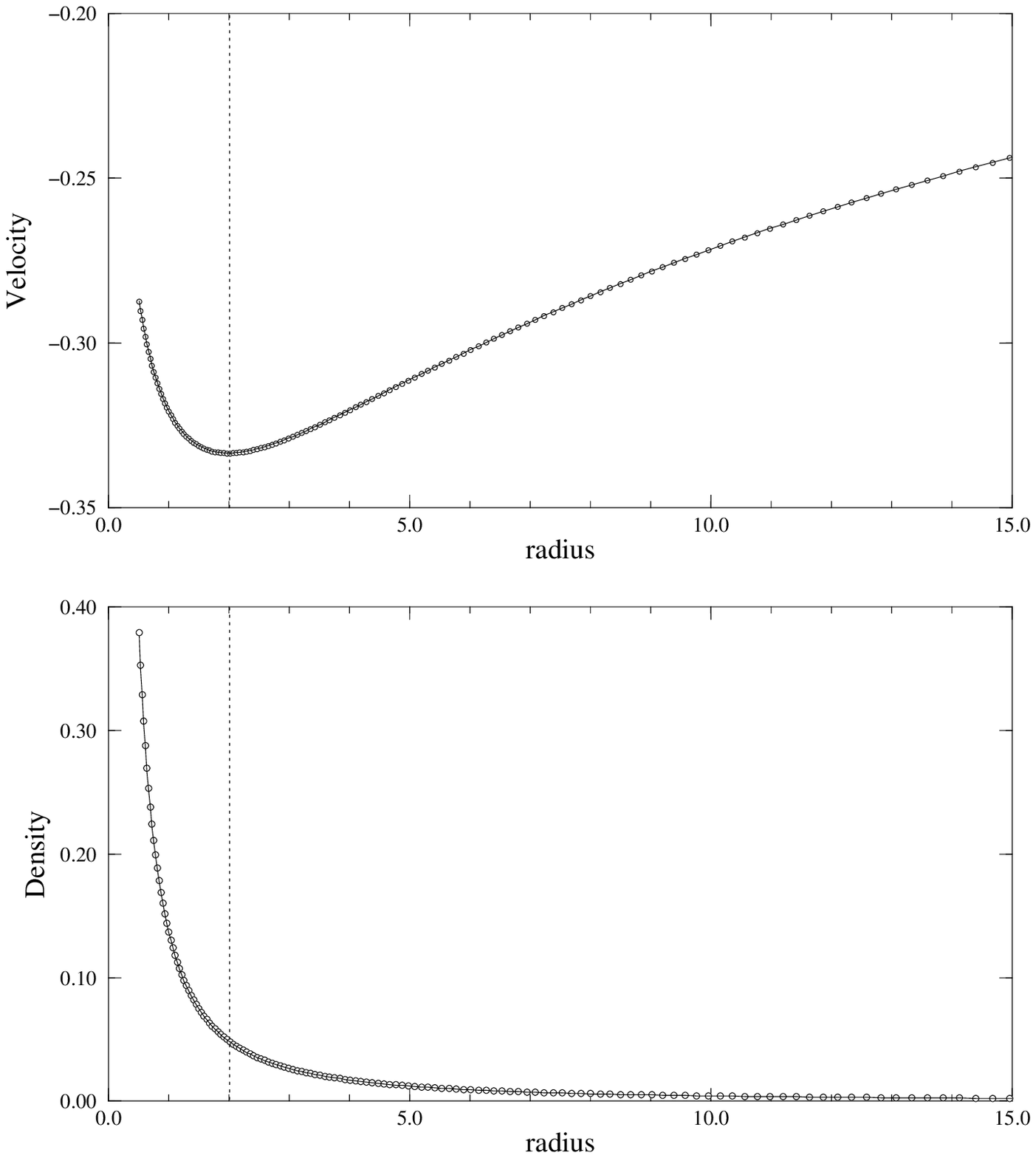,height=7.765in,width=6.0in}}
\caption{Exact (solid line) versus numerical (circles) solution for
spherical accretion of dust, after reaching the steady-state.  The top
curve shows the total velocity and the bottom one shows the rest-mass
density, as a function of the radial coordinate. The total domain
extends to $50M$. We focus on the first $15M$.  The value of
$C_1$ in Eq.~$(\ref{dens})$ was chosen to be $-0.195$. The numerical
solution agrees well with the exact one. The dotted line marks the
position of the horizon.}
\label{fig:1d-dust}
\end{figure}

\newpage

% FIGURE fig2.ps

\begin{figure}
\centerline{\psfig{figure=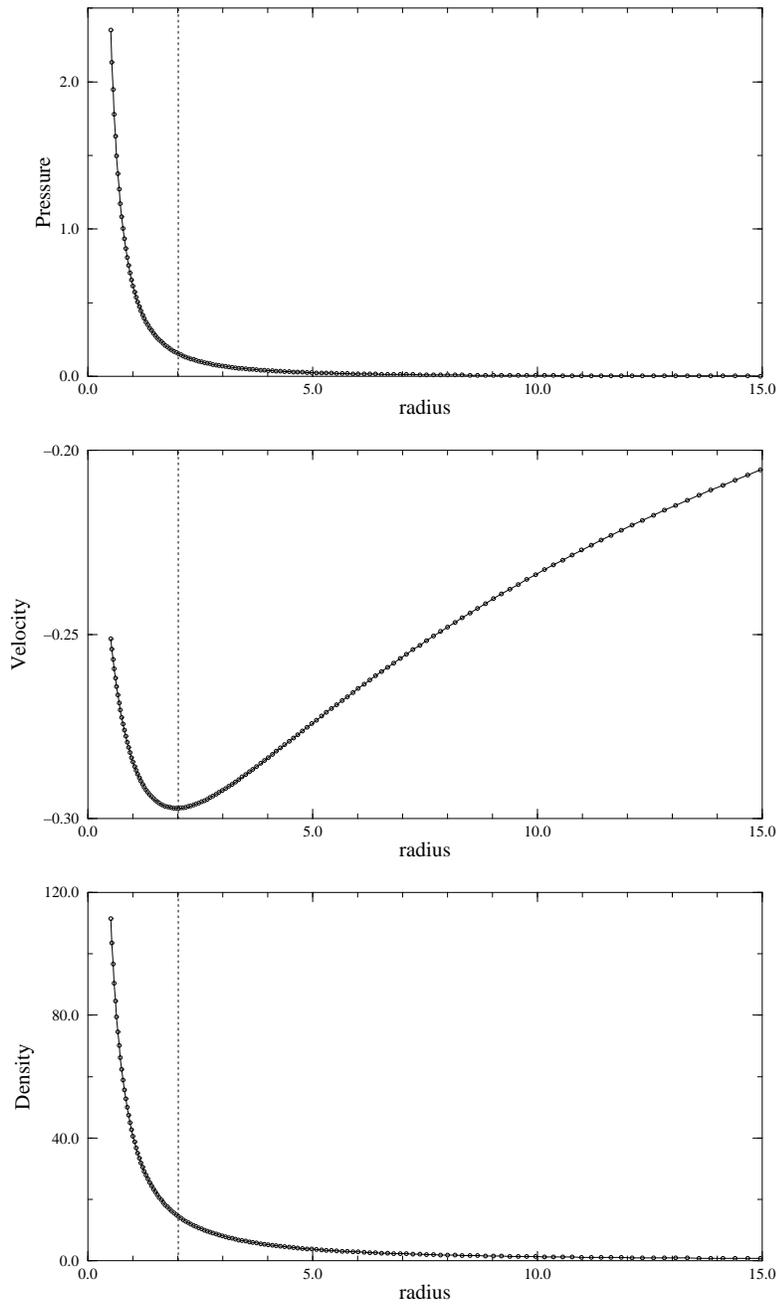,height=7.765in,width=6.0in}}
\caption{Exact (solid line) versus numerical (circles) solution for
spherical accretion of a perfect fluid, after reaching the
steady-state.  The top curve shows the pressure, the middle one shows
the total velocity and the bottom one shows the rest-mass density, as
a function of the radial coordinate.  As in Fig. 1, the total domain
extends to $50M$ and we only plot the first $15M$. The numerical and
exact solutions also match very well. The dotted line marks the
position of the horizon.  The critical point values used to construct
this particular exact solution are $r_c=400M$ and $\rho_c=10^{-2}$.}
\label{fig:1d-fluid}
\end{figure}

\newpage

% FIGURE fig3.ps

\begin{figure}
\centerline{\psfig{figure=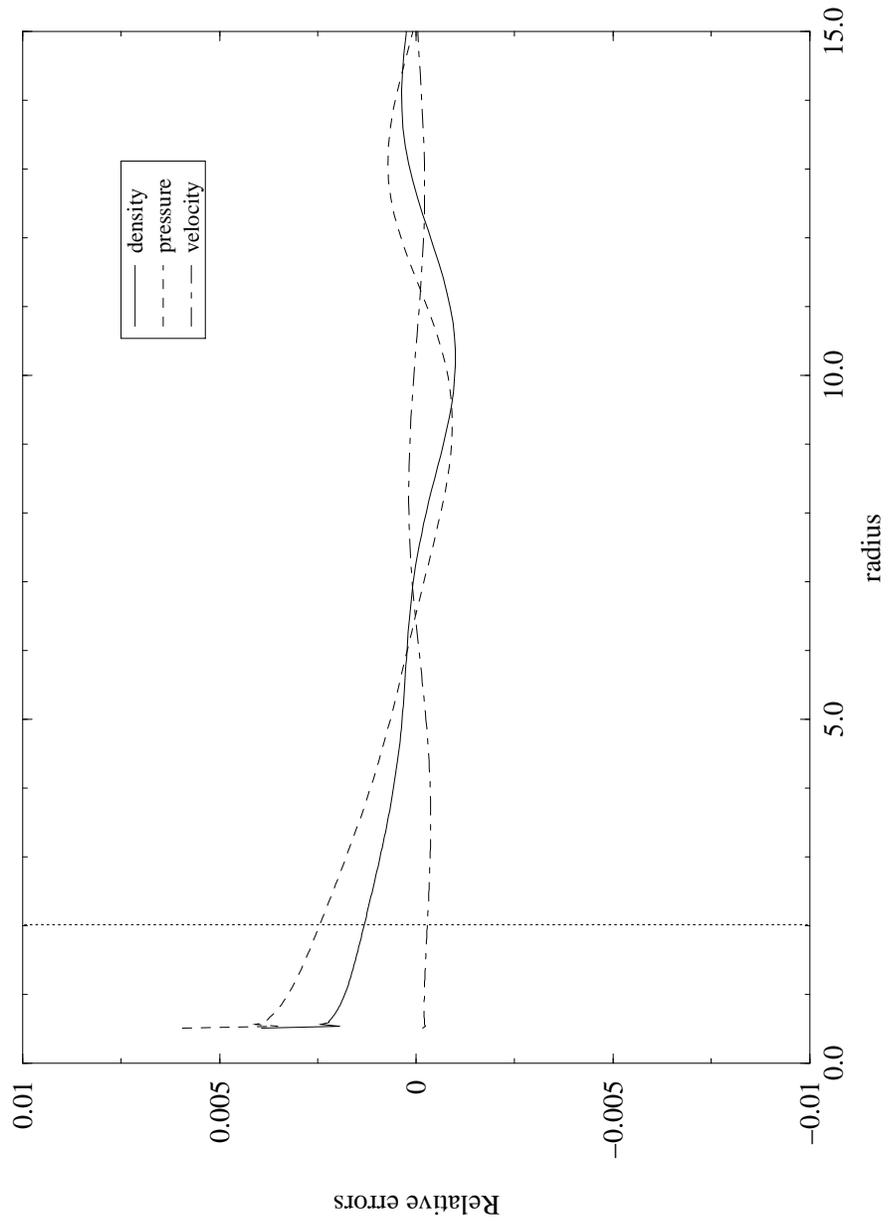,height=7.765in,width=6.0in}}
\caption{Relative errors for
density, pressure and total velocity in the case of perfect fluid
accretion, as a function of the radial coordinate $r$. The result is
for 200 grid points between 0.5 and $50M$. Only the first $15M$ are
shown. The overall good accuracy is apparent, the especially small
error in total velocity being most striking.}
\label{fig:rel_error}
\end{figure}

\newpage

% FIGURE fig4.ps
\begin{figure}
\centerline{\psfig{figure=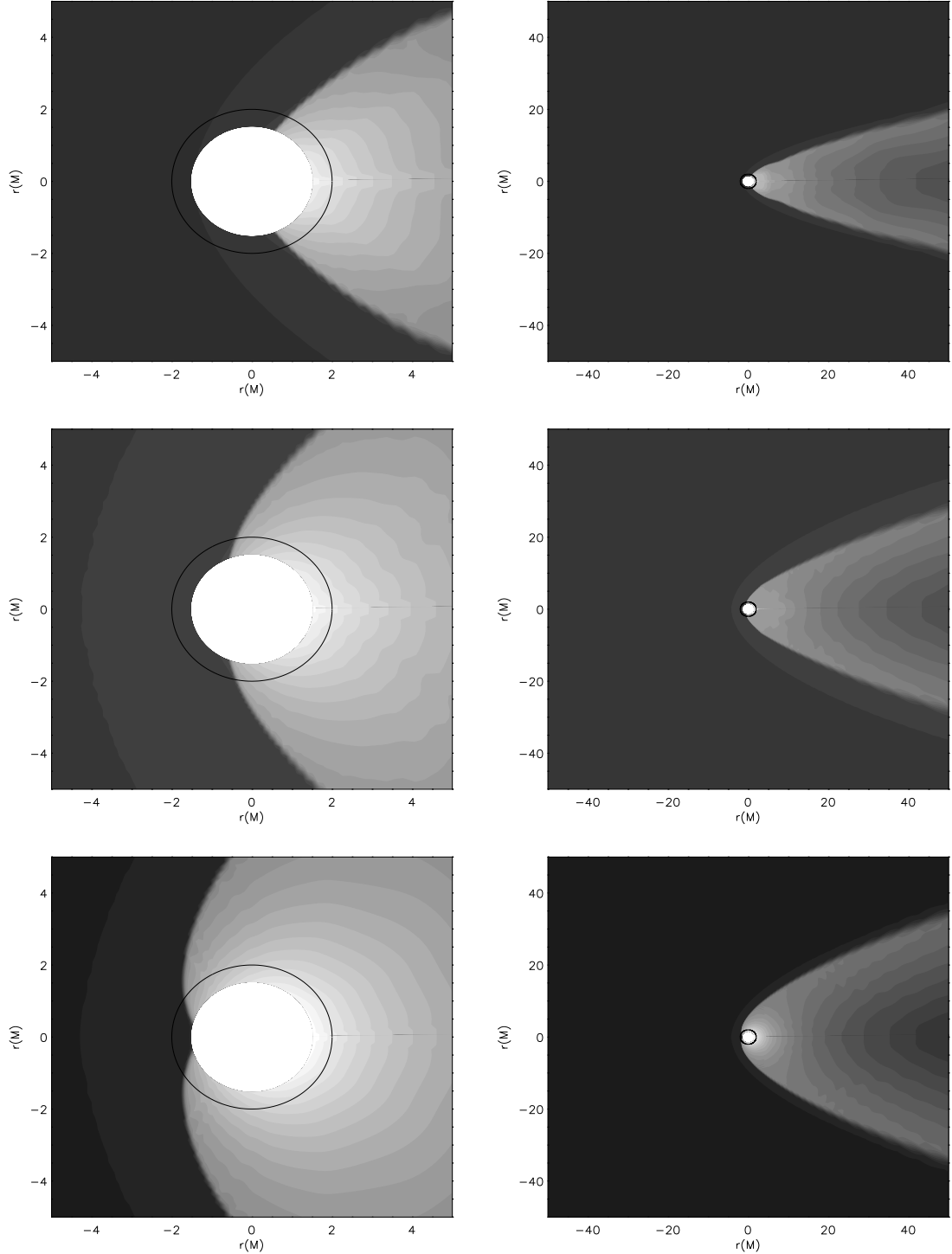,height=7.765in,width=6.0in}}
\caption{Grey-scale iso-contour plots of the logarithm of the 
rest-mass density for the axisymmetric relativistic Bondi-Hoyle
accretion problem. The final stationary solution is plotted. From top
to bottom we display the solution for $\Gamma=4/3$, $5/3$ and $2$,
respectively.  The left column shows a close-up view of the central
region, which is enlarged in the right column. The inner circle shows
the location of the horizon. The shock cone is well resolved even at
large distances from the hole. Its innermost location differs in the
three cases. For larger values of $\Gamma$ it moves to the front part
of the hole, due to the higher pressure values at its rear part.}
\label{fig:2d-bh}
\end{figure}

\end{document}